\documentclass[preprint, 3p, authoryear]{elsarticle}
\usepackage{algorithm}
\usepackage{amsmath}
\usepackage{amssymb}
\usepackage{amsthm}
\usepackage{hyperref}
\usepackage{xcolor}
\journal{Ocean Modelling}
\begin{document}
	
\begin{frontmatter}

\title{Linking Horizontal and Vertical Submesoscale Lagrangian Dynamics
	South of the Ibiza Channel}

% \author[affil]{given_name}{surname}

\author[1]{Sara Cloux \corref{corr}}{}
\ead{saracloux@gmail.com} %% correspondence author
\author[2,3]{Ismael Hernández-Carrasco}{}
\author[2]{Alejandro Orfila}{}
\author[1]{Emilio Hernández-García}{}
\author[1]{Cristóbal López}{}

\cortext[corr]{Corresponding author}

\affiliation[1]{
	organization={Institute for Cross Disciplinary and Complex Systems (IFISC), CSIC-UIB},
	% addressline={},
	% postcode={},
	city={Palma de Mallorca},
	country={Spain}
    }
\affiliation[2]{
	organization={Mediterranean Institute for Advanced Studies (IMEDEA), CSIC-UIB},
	% addressline={},
	% postcode={},
	city={Mallorca},
	country={Spain}
	}
\affiliation[3]{
	organization={Instituto de Ciencias Marinas de Andalucia (ICMAN-CSIC)},
	% addressline={},
	% postcode={},
	city={Puerto Real},
	country={Spain}
}

\begin{abstract}
Fronts, eddies, jets, and meanders populate the ocean surface and play a key role in organizing transport and mixing across a wide range of scales. From a Lagrangian perspective, frontal structures are conveniently detected as surface filaments that act as barriers to horizontal transport. However, the degree to which these structures maintain coherence below the surface and the relationship between horizontal and vertical motions in their vicinity remain poorly understood. Here we analyze two such filaments in the western Mediterranean Sea, south of the Ibiza Channel, using a high‑resolution ocean simulation and Lagrangian diagnostics specifically designed to characterize submesoscale transport. Backward Finite‑Size Lyapunov Exponents (FSLE) computed at successive depths reveal nearly vertical, curtain‑like structures extending several tens of meters into the water column, with 
intensity decreasing with depth. Vertical-velocity sections across the filaments show coherent downward motion aligned with the filament cores, while Lagrangian trajectories indicate rapid horizontal convergence of fluid particles accompanied by sustained downward displacement. The barrier effect of Lagrangian Coherent Structures (LCSs) is commonly associated with persistent dynamical features of large-scale flows. Here, however, we demonstrate that FSLE filaments generated by rapidly fluctuating submesoscale dynamics can also act as effective transport barriers. 
%This vertical transport is not confined to the filament core but spans a broader surrounding region. 
Taken together, these analyses reveal a clear coupling between horizontal convergence and vertical motion in FSLE filaments, consistent with subduction‑like processes, and highlight the interplay between mesoscale and submesoscale transport. The results demonstrate that the surface signatures of Lagrangian filaments identified through FSLE techniques correspond to three-dimensional transport structures that regulate both lateral aggregation and vertical connectivity in the upper ocean.
\end{abstract}

\begin{keyword}
	Lagrangian Diagnostics \sep Finite-Size Lyapunov Exponents \sep Mediterranean Sea \sep Coherent Structures \sep Submesoscale \sep Vertical Dynamics
\end{keyword}

\end{frontmatter}

% ------------- START OF PAPER ITSELF -------------

\section{Introduction}
\label{introduction}

Understanding oceanic transport processes is fundamental to monitor and to predict the physical and biological dynamics of the marine environment. Ocean circulation is highly structured, with mesoscale and submesoscale features such as fronts, filaments, and eddies, which strongly regulate the horizontal and vertical transport of water masses, nutrients, and passive tracers \citep{oschlies1998eddy,mcgillicuddy2016,von2020,yang2024}. While the horizontal dynamics of these structures has been widely investigated using satellite observations, high frequency radar data, drifters and models through both Eulerian and Lagrangian diagnostics \citep{Isern2003,dovidio2004,waugh2006,Lacasce2008,BeronVera2008,huhn2012horizontal,gula2014,Vansebille2018,hernandez2018effect}, submesoscale dynamics is increasingly recognized as key driver of vertical exchanges, linking surface and subsurface layers through ageostrophic motions that influence ocean biochemical cycle by supplying nutrients into the sunlit layer \citep{Levy2018,Mahadevan2015} and carbon dioxide subduction \citep{omand2015eddy}. Advances in high-resolution numerical modeling and observations now make it possible to explicitly resolve submesoscale circulation \citep{mcwilliams2016submesoscale,TarryFrontal2021,MartiSolanaBalanced2026} and to examine their three-dimensional structure and impact on ocean transport.

A variety of diagnostic approaches has been developed to identify organizing structures in unsteady flows \citep{Boffetta2001,Wiggins2005,Mancho2006,Lacasce2008,Prants2017,huhn2012,perez2013}, in particular Lagrangian ones aimed at characterizing the so-called Lagrangian Coherent Structures (LCS) \citep{Haller2015}. Among them, filamentary locations at which intense stretching or trajectory attraction occurs act as transport barriers and pathways in geophysical flows. The finite-size Lyapunov exponent (FSLE) is a Lagrangian technique that has been widely used as a practical diagnostic to highlight these locations of intense stretching or attraction  \citep{Boffetta2001,dovidio2004,hernandez2011reliable,cencini2013finite}, particularly in observational and model-based oceanographic studies. They are a convenient tool to identify frontal structures, identified as LCSs,  separating waters of different origins, and with a different future fate \citep{hernandezlagrangian,hernandez2018role}. FSLEs are usually computed at the ocean surface, where the identified frontal structures appear as filaments of large FSLE values. Studies at mesoscales have shown that LCSs that appear as filaments at surface remain approximately coherent for a range of depths, so that the whole three-dimensional structures consist of nearly-vertical curtain-like objects \citep{Branicki2010,Branicki2011,Bettencourt2012,bettencourt2013,Bettencourt2017}. 
The underlying reason is that vertical velocities in the ocean are much weaker than horizontal ones. As a result, the directions of strongest attraction and repulsion at any given point are expected to lie almost entirely within the horizontal plane. When this is combined with the incompressibility of the flow, it follows that the most attracting and repelling structures (i.e. the LCSs) should manifest as nearly vertical surfaces, since attraction or repulsion occurs normal to the LCS. Departures from verticality arise either from slight tilts of the dominant attracting or repelling directions away from the horizontal, or from strong vertical shear that induces vertical variations of those directions \citep{Branicki2011}.

Recent studies using high-resolution numerical simulations have revealed intense downward transport within submesoscale fronts undergoing frontogenesis \citep{verma2021lagrangian,capo2022coherent}. These events are associated with strong surface convergence and enhanced vertical velocities \citep{yu2024intensification}, while turbulence has been shown to dominate the variability of vertical motions within finite-scale regions of strong Lagrangian convergence \citep{capo2022coherent,aravind2023lagrangian}. Despite these advances, the relationship between such vertical transport processes and the surface signatures of attracting Lagrangian coherent structures, as identified from FSLEs, remains poorly understood. 

In this paper, we aim to clarify the extent to which the richer vertical dynamics present at submesoscales modifies the previous results, specifically how FSLE-identified surface attracting filaments exhibit vertical coherence and whether they constrain the three-dimensional motion of fluid particles. We use ocean velocities from a realistic simulation at submesoscale resolution in a region of the Western Mediterranean, from which we identify two particular FSLE-filaments. They will be taken as case studies to investigate the three-dimensional transport dynamics in their vicinities. 
We will use two complementary approaches: a) transport analysis through the identification of the LCS from the two-dimensional dynamics (i.e., neglecting the vertical velocities) at different depths, and b) recalculation of these structures from the full three-dimensional dynamics, complemented with analysis of three-dimensional fluid-particle trajectories in their vicinity. 

More in detail: First, we perform Lagrangian backwards simulations with only the horizontal components of the velocity field, restricting the 
particle trajectories to purely horizontal motion in successive depth layers (particles are initially released in layers at different depths). From these trajectories, computed backwards-in-time, we calculated two-dimensional FSLE fields for different depths.
This approach allows us to identify, at each depth, horizontal lines --the \emph{filaments}-- strongly attracting close fluid particle trajectories in the horizontal, and evaluate their vertical coherence (i.e. how they change vertically), showing that such structures are detectable to depths of around $50$ meters. We also confirm that, taken together, they form a three-dimensional LCS of  nearly-vertical curtain-like character.  Thorough the paper we will speak loosely of the \emph{position of the filament}, and \emph{proximity to the filament} when referring to the position and proximity to the FSLE filament identified at each depth, which maintains essentially the position of the surface filament. 
%%%%%%%%%%%%%%%%%%%%%%%%%%%%%%%%%%%%%%%%%%%%%%
Second, we allow the particles to evolve both backward and forward in time in three dimensions to perform fully three-dimensional Lagrangian numerical experiments. We recalculate the FSLE filaments under this richer three-dimensional dynamics (obtaining structures very similar to the ones given by the simpler 2D approach), analyze three-dimensional Lagrangian trajectories in their vicinity, and examine vertical velocity profiles. By studying the distance of particles to the filament core at each depth, we assess whether their vertical motions, including subduction processes and vertical modulations, occur under the dynamical influence of the coherent structure. Though our study is devoted to two particular FSLE-identified filaments, in a specified oceanic area at a given period of the year, we believe that our results provide a  generalizable  view of how surface filamentary structures organize transport in the upper ocean layers, showing the interaction between 
the horizontal constraints imposed by LCS and coherent vertical movements.

The paper is organized as follows. Section \ref{Sec:methods} describes the study area, the numerical model, and the methodologies used in this work. Section \ref{Sec:results} presents the results, including the identification of the surface filaments, their vertical coherence, associated vertical velocity structures, and three-dimensional particle dynamics. Finally, Section \ref{Sec:summary} summarizes the main conclusions and discusses the implications for three-dimensional ocean transport.

\begin{figure}[H] 
	\centering
	\includegraphics[width=\linewidth]{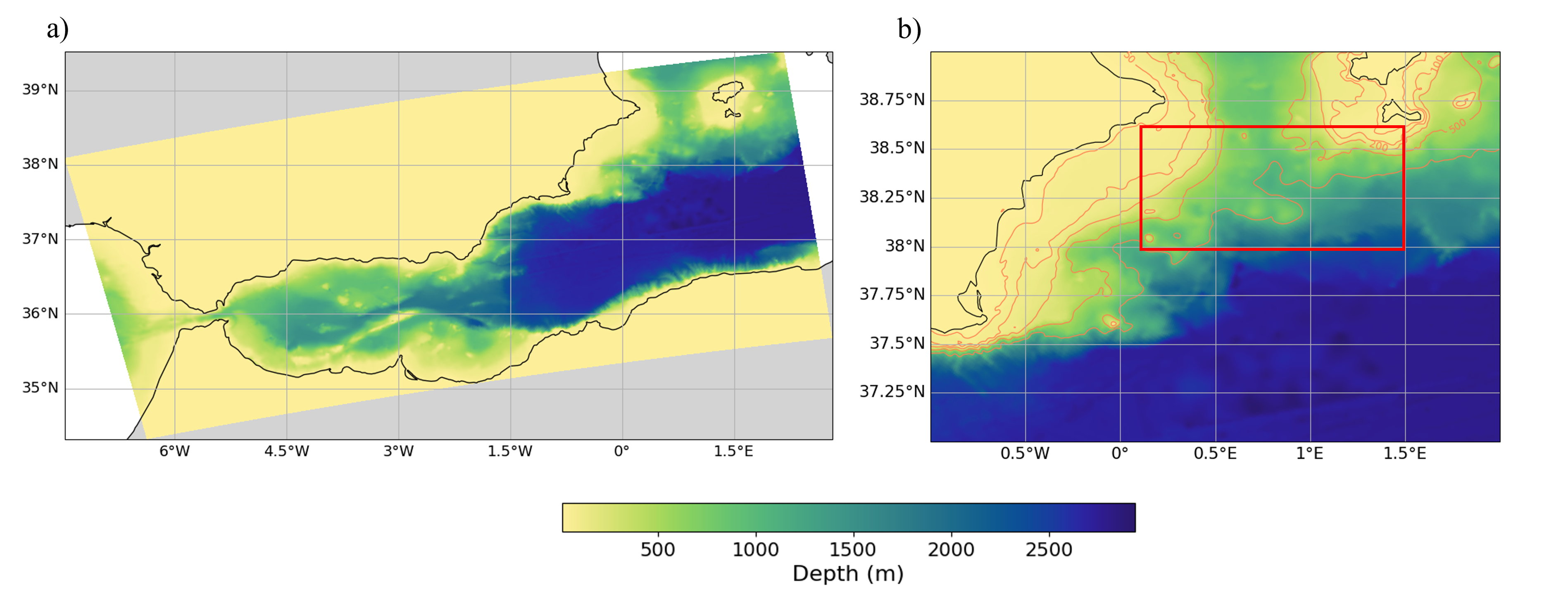}
	\caption{Model domain and study area. The background colour shading represents bathymetric depth (m) for (a) the CROCO simulation domain covering the Western Mediterranean Sea and (b) a zoomed-in view of the region south of the Ibiza Channel. The particle-analysis subdomain (38.0°–38.6°N, 0.1°E–1.5°E), where the Lagrangian diagnostics presented in this work are computed, is indicated by a rectangle in panel (b). Contour lines denote selected isobaths.}
	
	\label{fig:study_area}
\end{figure}

\section{Methods}
\label{Sec:methods}
\subsection{Study Area}

The study region, located in a transition zone between the Ibiza Channel and the Alboran Sea, connects the Western Mediterranean Sea with the Alboran Sea itself, covering an active corridor of the Western Mediterranean circulation (Figure \ref{fig:study_area}). This region (shown in Figure \ref{fig:study_area}b) extends approximately from 37.0°N to 39.0°N and from 0.2°W to 2.0°E, encompassing an active corridor of the Western Mediterranean circulation, an ecological "choke point" where, at surface, fresh Atlantic waters entering through the Strait of Gibraltar interact with the denser resident Mediterranean waters \citep{heslop2012autonomous}. Within this region, particle-based analyses are performed in a smaller subdomain located south of the Ibiza Channel, approximately between 38.0°N and 38.6°N and 0.1°E and 1.5°E. There, strong density gradients, shear, and topographic constraints give rise to pronounced mesoscale and submesoscale variability. Within this corridor, the circulation is characterized by the development of jets, filaments, and frontal structures that modulate cross-frontal exchange and enhance vertical motions, making the region particularly relevant for studying three-dimensional transport and dynamics of LCSs.

Surface frontal zones frequently occur near the peripheries of the gyres. These frontal systems, objective of this work, are characterized by strong  temperature and salinity gradients, and are often places where the rate at which submesoscale buoyancy gradients intensify driving the development of strong vertical motions \citep{allen2001,tintore1988}. 

\subsection{Input data}
\label{Sec:data}

The three-dimensional velocity fields are obtained from a high-resolution ocean simulation described in detail in \citet{Capo2021}. It has been performed using an offline grid-nesting procedure \citep{mason2010procedures} with increasing resolution using CROCO (Coastal and Regional Ocean Community model), a coastal ocean modeling system based on ROMS with a numerical kernel close to the UCLA version of the ROMS model (http://www.croco-ocean.org; Shchepetkin; McWilliams, 2009). The model domain covers a region larger than our study area, including the Alboran Sea and adjacent regions such as the Strait of Gibraltar, as shown Fig. \ref{fig:study_area}a), with a horizontal resolution of $500$ m. It extends from roughly 34°N to 38.62°N in latitude and from 6.5°W to 2.5°E in longitude. The vertical grid consists of $70$ sigma levels that follow the terrain, with an increasing number of layers, and therefore the resolution, in the surface and in the bottom, so that it allows an improved representation of processes associated with boundary layers, flow-topography interaction, and submesoscale dynamics,which make it ideal for studying vertical exchanges in the mixing layer. In this nesting approach, the parent grid that provides initial and boundary conditions to the finer child grid, covers the whole western Mediterranean Sea and part of the Atlantic Ocean, with a horizontal resolution of 1.5 km and 60 sigma levels. Inital and boundary conditions are forced with daily averaged fields of temperature, salinity, currents and sea level height interpolated from the Mercator Global Ocean forecasting product at 1/12º horizontal resolution distributed by CMEMS, and hourly atmospheric data at 0.2º from the NCEP-CFSR product. Tidal forcing is introduced using a global tidal atlas, specifically the OSU-TOPEX/Poseidon Global Inverse solution database (version TPXO.7), while bathymetric information is taken from the SRTM30\_PLUS\_V11 dataset \citep{becker2009}. Model outputs are stored at three-hour intervals, providing a full year of temporally resolved three-dimensional velocity fields suitable for computing Lagrangian trajectories. These velocity fields are then used to drive offline particle-tracking experiments.  

\subsection{Lagrangian methods}
%\label{Sec:Lyapunov}

Our approach is based on a Lagrangian analysis of marine flows, in which particles are advected by the three-dimensional velocity field extracted from the high-resolution numerical simulation described in the previous section. The trajectories are computed by numerically integrating the transport equations below, Eq. (\ref{eq:advection}). For every numerical experiment, particles are initially distributed homogeneously throughout the study region and at multiple depths. Positions are expressed in geographical coordinates (longitude $\lambda$, latitude $\phi$, and depth $z$), and the advection equations account for the Earth's spherical geometry:

\begin{equation}
	\label{eq:advection}
	\left(\frac{d\lambda}{dt},\frac{d\phi}{dt},\frac{dz}{dt}\right) = 
	\left(\frac{u(\lambda,\phi,z,t)}{R \cos\phi}, 
	\frac{v(\lambda,\phi,z,t)}{R}, 
	w(\lambda,\phi,z,t)\right),
\end{equation}
where $R$ is the Earth's radius, and $(u,v,w)$ are the zonal, meridional, and vertical velocity components. The velocity field is interpolated in space and time from the discrete numerical model outputs using a multilinear interpolation scheme, providing continuous 
velocity estimates required for trajectory integration.

Large values of FSLE highlight regions of enhanced stretching or contraction, thereby identifying LCSs embedded in the flow field \citep{Boffetta2001,dovidio2004,hernandez2011reliable,cencini2013finite}. The FSLE at a given location is defined proportionally to the inverse of the minimum time $\tau$ required for a reference particle, initially located at time $t_0$ at position ${\bf x}_0=(\lambda_0,\phi_0,z_0)$,
and its four horizontal nearest neighbors (initially separated by a distance $\delta_0$, measured along the spherical surface), 
to reach a final separation distance $\delta_f$. More specifically:
\begin{equation}
	\textrm{FSLE}({\bf x}_0,t_0,\delta_0, \delta_f) = \frac{1}{\tau} \log \frac{\delta_f}{\delta_0}.
	\label{eq:FSLE}
\end{equation}
If the integration is done forward in time, large values of FSLE identify initial locations ${\bf x}_0$ from which fluid parcels would experience large stretching, producing large particle separation rate. If integration is done backwards, the high-FSLE values identify locations of strong attraction for the trajectories. 

In this work we compute FSLEs backward in time since they identify attracting structures, which are among the most dynamically significant features of the flow \citep{dovidio2004}. The spatial scales effectively sampled by the FSLE are determined by the choice of the initial and final separation scales  $\delta_0$ and $\delta_f$. In this work we focus on submesoscale dynamics and thus consider the range $[\delta_0,\delta_f]$ to span scales of the order of few kilometers. In backward-time computations, the characteristic time $\tau$ represents the time required for a pair of particle trajectories to converge from a separation $\delta_f$ to a separation $\delta_0$, and the FSLE therefore quantifies the local convergence rate at the selected spatial scales.

As indicated before, we perform two complementary types of Lagrangian calculations:
\begin{enumerate}
	\item
	First, we compute backward-in-time particle trajectories constrained to purely two-dimensional motion on successive horizontal layers. Hereafter, the FSLE fields derived from these two-dimensional calculations are referred to as FSLE-2D. Particles are initialized on a regular horizontal grid with a spacing of approximately $\delta_0 = 250$ m and covering the study area, i.e. from latitude 38$^\circ$ N to 38.6$^\circ$ N and from longitude 0.1$^\circ$ E to 1.5$^\circ$ E (see Figure \ref{fig:study_area}b). This results in approximately $150.000$ particles per depth and per simulation. Particle ensembles are released every hour during the period from January 2011 to April 2011, and every $5$ meters in the vertical dimension from the surface to $50$ m of depth. For each release time $t_0$, backward-in-time trajectories are integrated over 15-day time windows at different depth levels using only the horizontal velocity components ($w=0$ in Eq. (\ref{eq:advection})). Positions are monitored every 1 hour, allowing us to compute backward horizontal FSLE fields. We choose as the final horizontal separation distance $\delta_f = 5$ km. This enables the FSLEs to characterize horizontal dispersion across (submesoscale) spatial scales ranging from $250$ m to $5$ km. 
	
	\item
	In a second set of simulations, the same experimental configuration is repeated but using the full three-dimensional velocity field. The FSLE fields obtained from these three-dimensional calculations are hereafter referred to as FSLE-3D. All release locations, frequencies, and integration windows are kept unchanged, but particles are allowed to move in three dimensions. Backward-in-time trajectories are first computed to recalculate the FSLE fields including the effects of vertical motion and to assess whether the resulting Lagrangian structures remain consistent with those obtained from the 2D calculations. In addition, forward-in-time trajectories are integrated to investigate transport pathways, connectivity patterns, and the influence of vertical motions on particle dispersal. Apart from the inclusion of the vertical velocity component, all other aspects of the experimental setup remain unchanged.

\end{enumerate}

This combined analysis allows us to establish a direct connection between the horizontal organization revealed by the maximum values of FSLEs and the vertical dynamics experienced by the particles close to the coherent structures. To further quantify the interaction between particles and the LCSs identified by the large-FSLE structures, we compute the horizontal distance between each particle and the corresponding large-FSLE filament. In principle the distance should be done for each particle position with the corresponding filament obtained at the same depth. But we will see that the filaments obtained at different depths are aligned in the vertical forming nearly-vertical curtains. Then it is enough to compute the distance from the horizontal coordinates of particle position and the surface filament. More explicitly, the minimum distance between the particle positions and all grid points belonging to the surface filament is calculated at each time step. Distances are computed on the sphere using the haversine formulation, assuming an Earth radius of $R = 6371$ km. This procedure yields a time-dependent distance for each particle, which characterizes the convergence toward the filament, residence times in its vicinity, and subsequent dispersion. Next section shows the results.

\begin{figure}[h] 
	\centering
	\includegraphics[width=\linewidth]{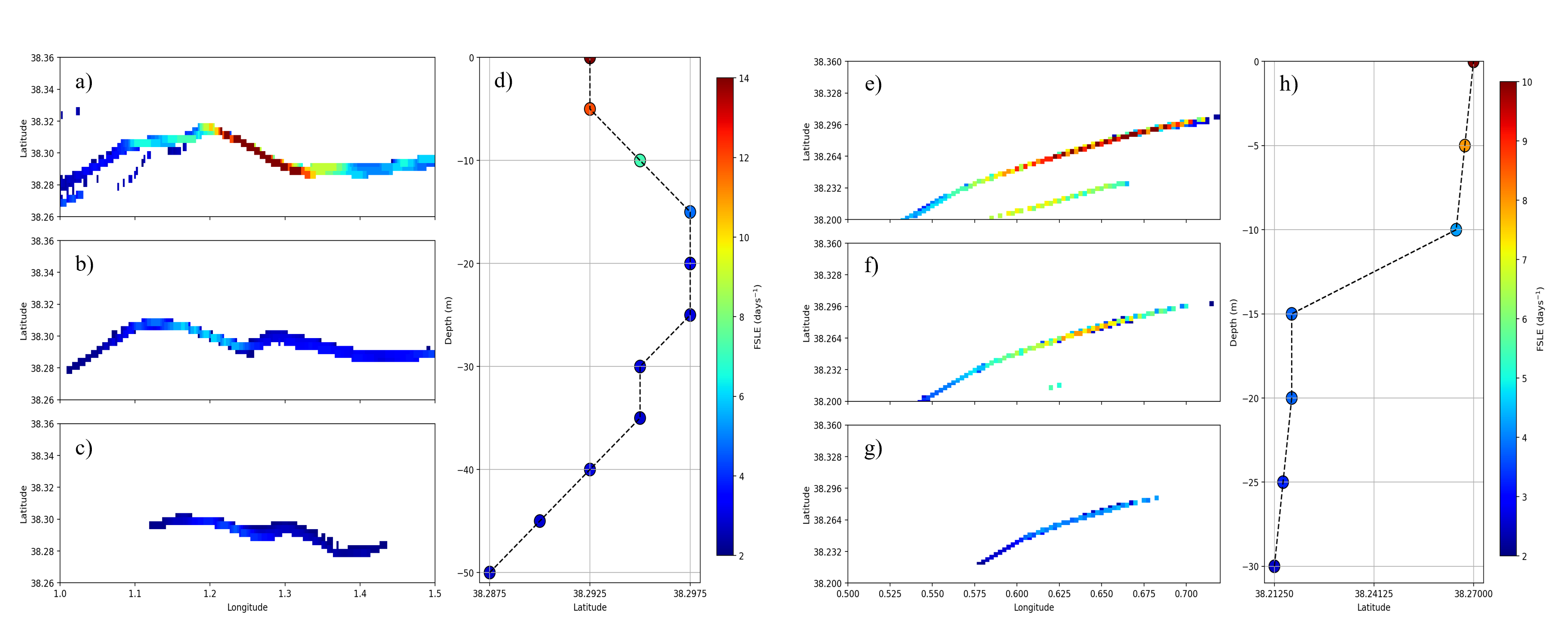}
	\caption{Horizontal and vertical distribution of backward FSLE-2D as a function of the initial release location ${\bf x}_0$ during two events: release time $t_0$ on February 27 at 18:30 (left group of panels) and release time $t_0$ on March 20 at 9:30 (right group of panels). For the first event, panels (a)--(c) show the horizontal FSLE-2D structure at depths of 0, $-25$, and $-45$~m, respectively, while panel (d) shows the corresponding vertical distribution. For the second event, panels (e)--(g) show the horizontal FSLE-2D structure at depths of 0, $-10$, and $-20$~m, respectively, while panel (h) shows the corresponding vertical distribution. Only locations with $FSLE-2D > 2~\mathrm{day}^{-1}$, identifying the filaments, are displayed in color; regions with lower FSLE-2D values are shown in white. Panels (d) and (h) show, for each horizontal layer at 5~m depth intervals, the latitudinal coordinate of the maximum FSLE-2D value within the domain displayed in the corresponding horizontal panels.
	}
	\label{fig:fsle_in_depht}
\end{figure}

\section{Results and Discussion}\label{Sec:results}

Among the regions and date ranges in which attracting LCSs in the form of FSLE filaments at different depths have been identified, we analyze here two of them, which turn out to be representative of examples of the structures found. Figure \ref{fig:fsle_in_depht} illustrates the horizontal distribution of backwards-in-time FSLE-2D fields for the two representative structures at different depths, computed using purely two-dimensional particle trajectories at successive depths with $\delta_0=250$ m and $\delta_f = 5$ km. We define an \textit{FSLE-2D filament} as the region where FSLE-2D values exceed $2~day^{-1}$. This treshold value corresponds to mixing times of approximately 1,5 days, which are the typical times in this study. To emphasize these structures, only values above this threshold are shown in color in the figures, while the rest of the domain,  where FSLE-2D$<2~day^{-1}$, is masked in white (this threshold removes weaker FSLE-2D signals that are not associated with the filament core, allowing the filamentary structures to be more clearly distinguished from the background field). The structures in Figure \ref{fig:fsle_in_depht} are computed at the time $t_0$ at which each  filament takes its \textit{maximum intensity}. This time $t_0$ is defined as the time at which the average over all the horizontal area displayed in Figs. 2a and 2e of the FSLE-2D values (all values in the area, not just the ones above the threshold of $2~day^{-1}$) is the largest one within the temporal window explored. 

We note that both FSLE-2D filaments remain approximately at the same horizontal location across the different depths, although fading in intensity. To confirm in a more quantitative way this nearly-vertical curtain-like character of the high-FSLE-2D structures, panels d) and h) in Fig.  \ref{fig:fsle_in_depht} show a synthetic view of the organization with depth of each filament. These are constructed by identifying, at the surface, the location of the FSLE-2D maximum associated with the filament and then tracking it downwards by selecting, at each depth, the 
closest FSLE-2D maximum in horizontal distance.

The first filament (Figure \ref{fig:fsle_in_depht}a) extends (at surface) from longitudes 1–1.5$^\circ$E and latitudes 38.20–38.36$^\circ$N and remains detectable (i.e. above the $2~day^{-1}$ threshold) until depths of approximately 50 m. Its marked vertical coherence can be quantified by noticing that the largest latitudinal displacement of its maximum FSLE-2D-value is of about 0.01$^\circ$, corresponding to about 1 km, to be compared to the approximate length of the filament at surface, which is about 0.5$^\circ$ (close to 50 km). This confirms that this three-dimensional attracting structure is essentially vertical.  Its strongest signature occurs at surface, for $t_0$ given by 27 February at 18:30, when the peak of FSLE-2D values is about 14 days$^{-1}$. We note that, although the FSLE-2D fields have been computed by following (backwards) particle separation in the range $[\delta_0,\delta_f]=[250 m, 5 km]$, within submesoscales, the identified filament has a much larger length, implying that both mesoscale and submesoscale dynamics are involved in determining transport in the explored range. 

The second filament (Figure \ref{fig:fsle_in_depht}b), located at surface between longitudes 0.5–0.7$^\circ$E and latitudes 38.20– 38.30$^\circ$N, displays a significantly shallower vertical extent. Shown here at its peak intensity at $t_0$ on 20 March at 09:30, the filament is clearly identifiable at 0, 10, and 20 m depth, but becomes undetectable below approximately 30 m depth, where FSLE-2D values fall below the imposed threshold. This rapid vertical decay indicates that the associated attracting submesoscale structure is confined to the near-surface layers. The maximum latitudinal displacement across different depths is about 0.06$^\circ$, six times larger than for the first filament. Thus, although the resulting three-dimensional structure is still near to  vertical, it shows a weaker vertical coherence, probably linked to stronger influence of surface processes.

The comparison between both filaments shows distinct regimes of vertical organization. The first filament extends to a deep layer, whereas the second one is restricted to the upper tens of meters. In both cases, FSLE-2D values decrease with depth, reflecting a weakening of horizontal stirring. However, the sharper decay observed in panel (b) suggests a more stratified flow. These differences point to variability in the three-dimensional structure of mesoscale transport barriers, even when their surface signatures may appear qualitatively similar.

\begin{figure}[htb] 
	\centering
	\includegraphics[width=\linewidth]{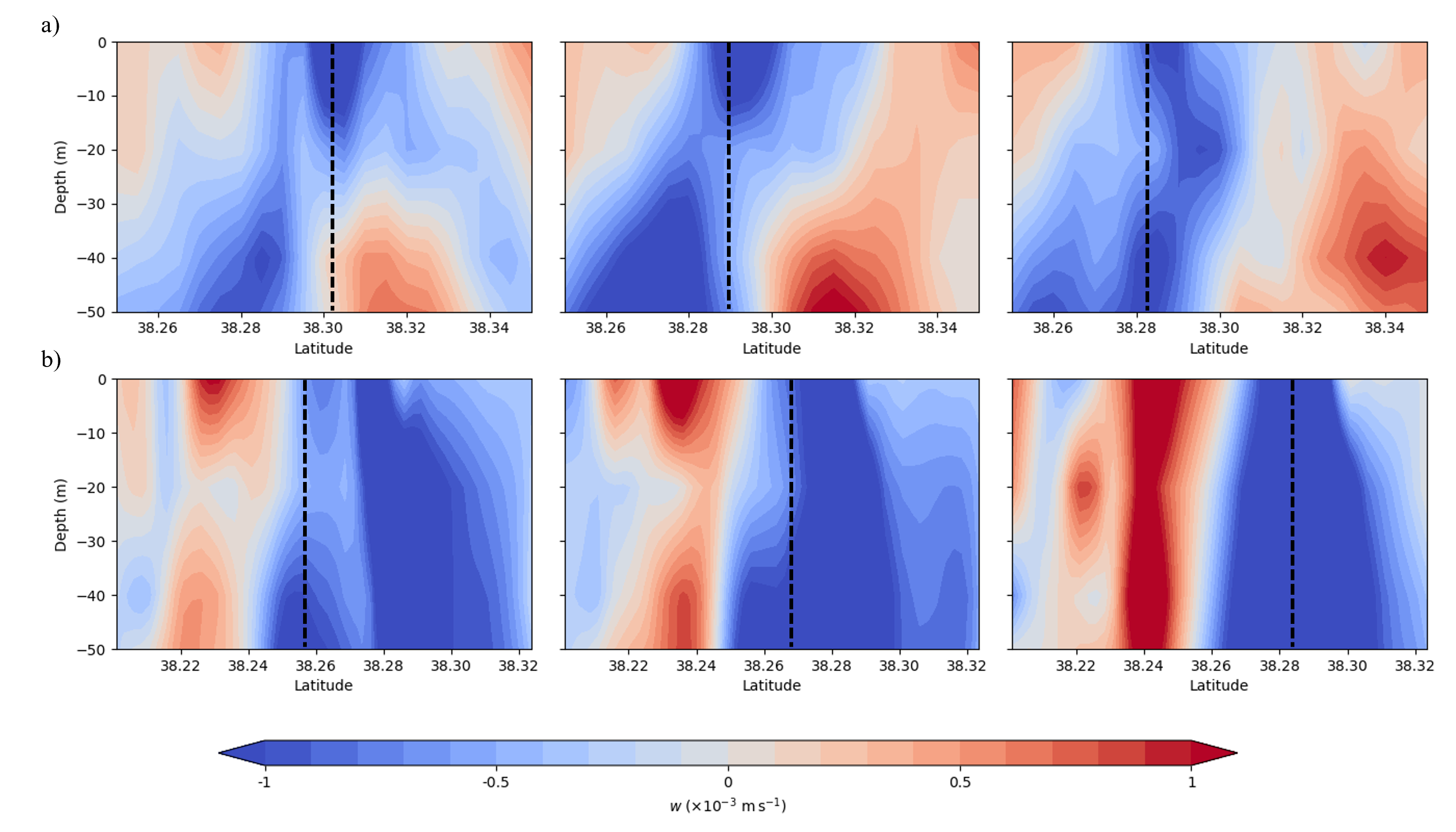}
	\caption{Values of the vertical velocities (m/s) on cross sections of the two filaments identifying in Fig. \ref{fig:fsle_in_depht}. 
		(a) Cross sections for the first 
		filament (1–1.5$^\circ$E, 38.20–38.36$^\circ$°N) 
		at longitudes (from left to right) 1.275$^\circ$E, 1.295$^\circ$E, 
		and 1.325$^\circ$E, at 27 February at 18:30, 
		and (b) cross sections 
		for the second filament (0.5–0.7$^\circ$E, 
		38.20–38.325$^\circ$N) at longitudes 
		0.60$^\circ$E, 0.62$^\circ$E, and 0.65$^\circ$E and 20 March at 09:30. The dashed line indicates the 
		latitude at which the corresponding filament is located at surface.}
	\label{fig:w_vel_cuts}
\end{figure}

Though our study is based on Lagrangian diagnostics, an Eulerian analysis may complement it. In Figure \ref{fig:w_vel_cuts} we show vertical cross-sections of the vertical velocity field, $w$, along the locations of the two considered filaments. The sections are plotted near the time of maximum filament intensity (27 February at 18:30 for the first event and 20 March at 09:30 for the second one). Panel (a) shows sections across the first filament at longitudes between 1–1.5$^\circ$E, while panel (b) presents analogous sections for the second filament between 0.5–0.7$^\circ$E. In both cases, a vertically-aligned coherent region of negative vertical velocity is observed from the surface down to approximately 40–50 m, with maximum intensity near the filament cores. These downward motions form narrow bands that are well aligned with the surface position of the filaments, as indicated by the dashed line. This spatial correspondence indicates that the FSLE-2D identified filaments are associated with coherent vertical motions, rather than being purely surface features. The coexistence of downward and upward motions suggests a frontal-like structure, in which horizontal particle trajectories convergence is linked to localized subduction. This is confirmed in the following by studying the full three-dimensional trajectories around the filaments.

\begin{figure}[H] 
	\centering
	\includegraphics[width=0.9\linewidth]{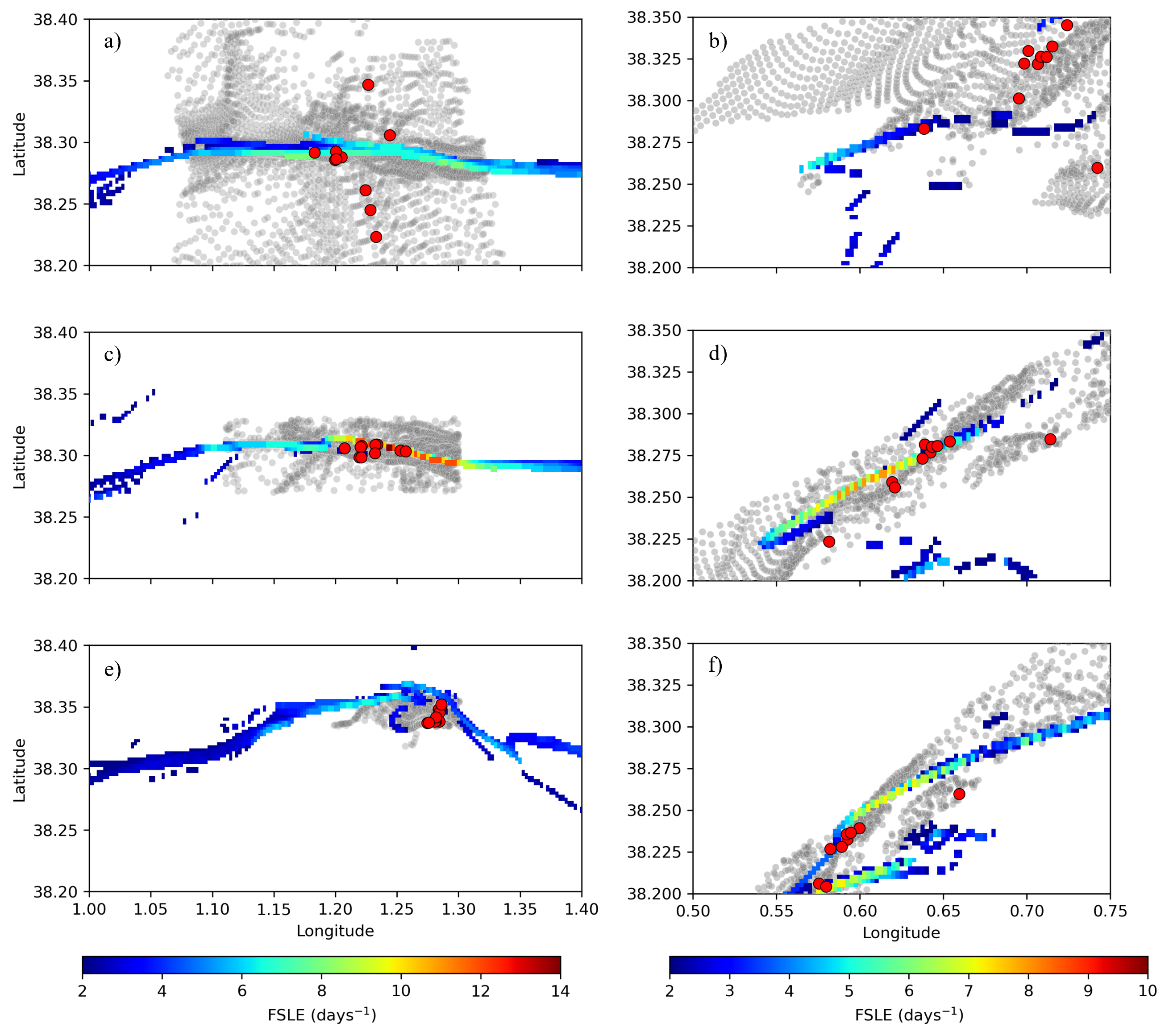}
	\caption{Temporal evolution of two FSLE-3D filament events and their interaction with Lagrangian particles (obtained from 3D trajectories). Colored pixels show backward FSLE-3D values at a depth of 10 m, computed from fully three-dimensional trajectories (only points with FSLE-3D values larger than  $2~\mathrm{day}^{-1}$ are displayed). Left column corresponds to the February event and right column to the March event. Panels (a), (c) and (e) correspond to 27 February at 06:30, 18:30 and 28 February at 21:30, respectively, for the time $t_0$ of backwards trajectory release, whereas panels (b), (d) and (f) correspond to 19 March at 21:30, 20 March at 09:30 and 20 March at 18:30, respectively. The central panel for each event (i.e. panels c) and d)) correspond the time of maximum intensity of the filament. Grey markers indicate the horizontal coordinates of a subset particles that were released in a large area at 25 m depth approximately 10 h before the time of maximum FSLE-3D intensity and subsequently evolved in a full three-dimensional forward simulation. The particles were included in the plotted subset if a condition of proximity to the filament, at any depth, was satisfied at the time of maximum filament intensity (see main text). Ten representative particles are highlighted in red.}
	\label{fig:traj_y_fsle}
\end{figure}

We now describe the results from Lagrangian simulations that use the full 3D velocity field. Particles were initially released at different depths and times, as in the 2D simulations, but now they evolve under the three velocity components $(u,v,w)$. FSLE fields were again computed in the region of interest, and at dates close to the relevant ones for the filaments identified in the 2D study. Structures in the FSLE-3D are very similar to the ones found in the FSLE-2D analysis, with the same vertical curtain-like structure, intensity fading with depth, and nearly identical locations and time of maximum intensity, although some minor differences are present. Figure \ref{fig:traj_y_fsle} shows the temporal evolution of the two filament events, recomputed now from the 3D trajectories with an initial release depth of 10 m. The central panels (c and d) were computed for the release time $t_0$ for which the filaments attain their maximum intensity (which correspond to the same time of maximum intensity for the 2D February filament shown in Fig. \ref{fig:fsle_in_depht}). Again, for the 3D filaments we define the time of maximum intensity as the value of $t_0$ for which the average of all FSLE-3D values inside the area plotted in Fig. \ref{fig:traj_y_fsle} is the largest one in the time window explored. Snapshots before and after this reference time (always with particles release at 10 m depth) are shown to illustrate the evolution of the structures along their life cycle. For the first event, the selected dates are 27 February at 06:30 (Fig. \ref{fig:traj_y_fsle}a), 27 February at 18:30 (Fig. \ref{fig:traj_y_fsle}c), and 28 February at 21:30 (Fig. \ref{fig:traj_y_fsle}e). For the second event, the corresponding times are 
19 March at 21:30 (Fig. \ref{fig:traj_y_fsle}b), 20 March at 09:30 (Fig. \ref{fig:traj_y_fsle}d), and 20 March at 18:30 (Fig. \ref{fig:traj_y_fsle}f).  

\begin{figure}[H] 
	\centering
	\includegraphics[width=0.9\linewidth]{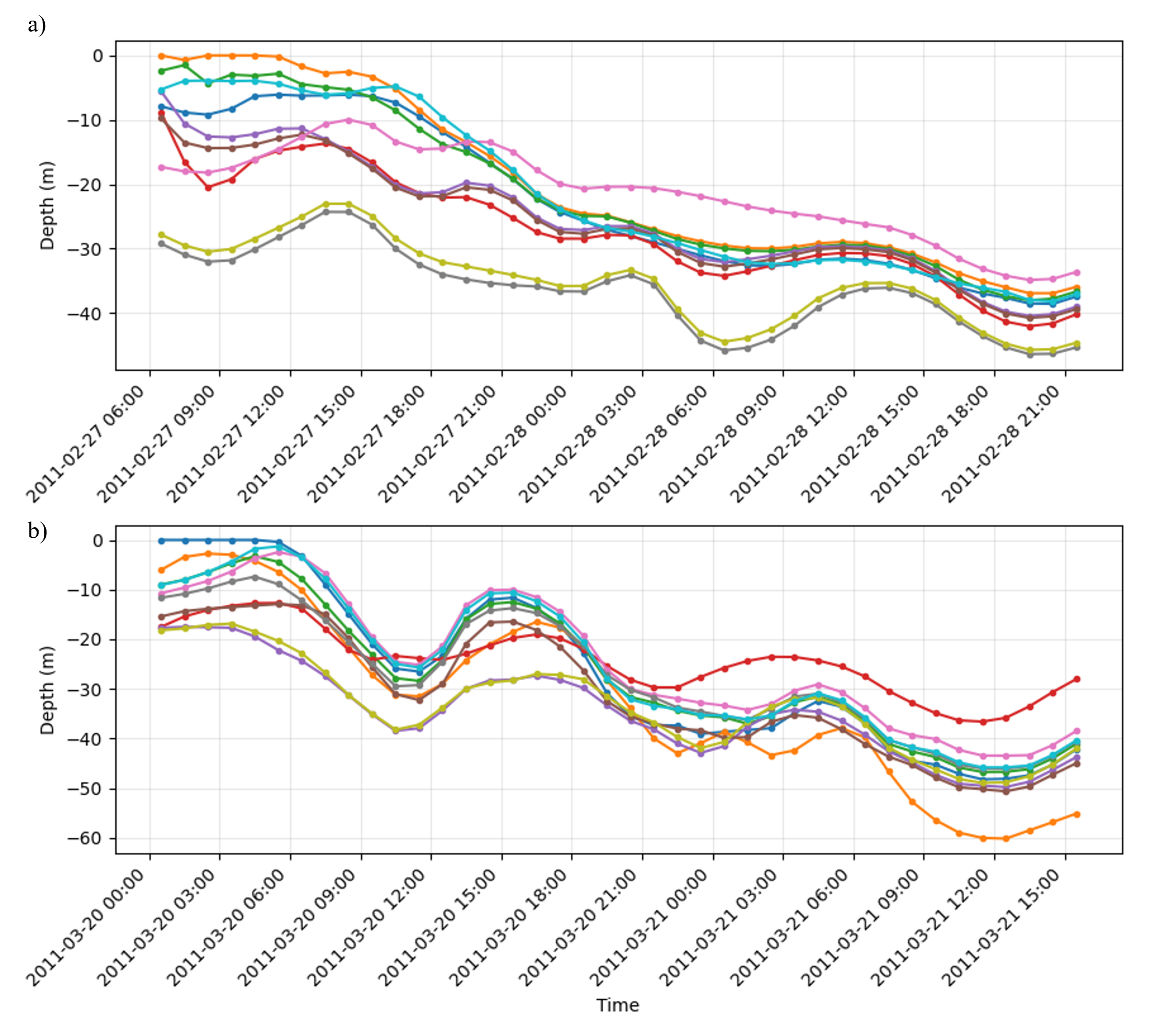}
	\caption{Vertical evolution in time of  
		particle trajectories for the same 
		particle groups shown in 
		Figure \ref{fig:traj_y_fsle}. Panel (a) corresponds to the first event, while panel (b) corresponds to the second event.}
	%as influenced by the local FSLE structure.
	\label{fig:traj_vertical}
\end{figure}

The filaments are shown in color in the foreground of Fig. \ref{fig:traj_y_fsle}. In the same figure, a set of particle positions are displayed in the background as semi-transparent grey markers (with ten of them highlighted in red). This set of displayed particle positions is obtained as follows:  For each event, particles were released (following the release strategy, particle number and forward-in-time trajectory configuration described in Section 2.3) at an intermediate depth of 25 m approximately 10 h before the time of maximum FSLE-3D intensity, allowing them to evolve under the flow field, so that they explore different horizontal and vertical locations. At the time of maximum filament intensity (panels c and d in Fig. \ref{fig:traj_y_fsle}) the subset of particles at any depth, within an approximate latitudinal band of $\pm 0.2^\circ$ around the filament and along the full longitudinal extent of its most intense sections were selected. This is the set of particles displayed in grey in all panels corresponding to each event in Fig. \ref{fig:traj_y_fsle}. We note that particle trajectories are fully three-dimensional, whereas only their horizontal coordinates are displayed in Fig. \ref{fig:traj_y_fsle}. However, because the attracting FSLE-3D structures analysed here exhibit a nearly vertical geometry, horizontal proximity of particles to the filament implies proximity to the corresponding FSLE-3D structure at depth. Consequently, the horizontal projections shown in Fig. \ref{fig:traj_y_fsle} provide a meaningful representation of the interaction between particles and the attracting LCSs throughout the water column.

%The FSLE filaments shown here were computed from backward-in-time trajectories evolving under the full three-dimensional velocity field. Consequently, their geometry differs slightly from the FSLE structures presented in Figure \ref{fig:fsle_in_depht}, which were obtained from trajectories constrained to horizontal motion. Nevertheless, both approaches identify attracting structures in the same geographical region and at essentially the same stages of the two events, indicating that the detected filaments are robust to the inclusion of vertical motion.

For the first event (left column), at 27 February at 06:30 the set of selected particles is distributed on both sides of the filament. As the filament approaches its peak at 27 February at 18:30, particle positions begin to converge and align with the structure. By 28 February at 21:30, the filament weakens and deforms, while the particles remain clustered along it, consistent with a backward-attracting LCS behavior. This evolution suggests a  three-stage process: an initial dispersed phase, a rapid focusing phase near the time of maximum FSLE-3D intensity, and a later stage in which particle clustering persists even as the filament weakens.

For the second event (right column), at 19 March at 21:30 the set of selected particles is located mostly away from the filament. At 20 March at 09:30, particles  align along the filament as it reaches maximum intensity, again showing convergence toward the FSLE-3D ridge. By 20 March at 18:30, the filament begins to lose intensity, but particles remain attached to it over time. Compared to the first event, in this case the filament appears more curved and fragmented, leading to a less uniform particle distribution along it. Particles are less strongly attracted to the FSLE-3D filament, sugesting a weaker convergence within the filament and, likely, shallower vertical penetration than that of the other filament.

\begin{figure}[htb] 
	\centering
	\includegraphics[width=0.9\linewidth]{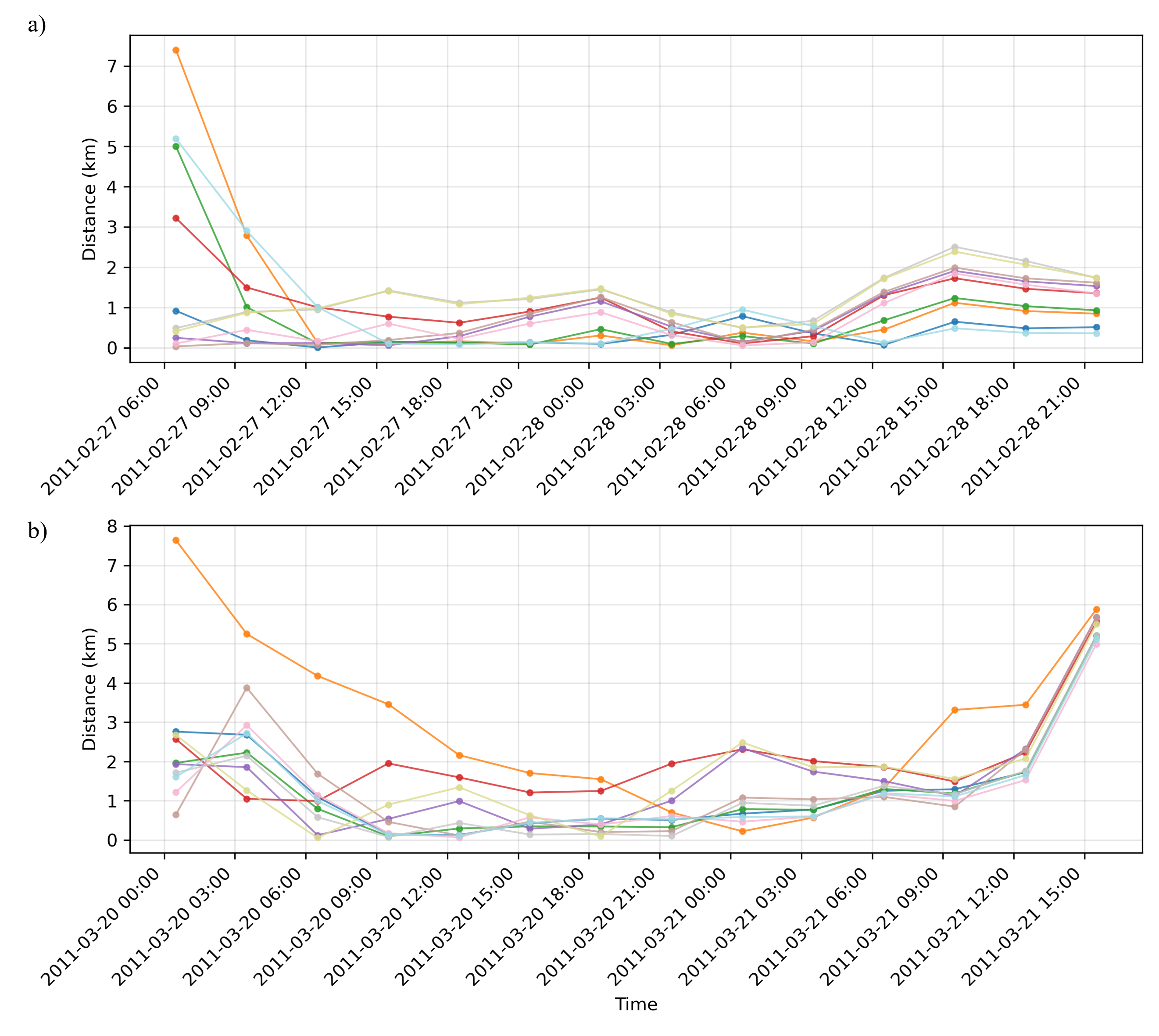}
	\caption{Temporal evolution of the horizontal distance between particles and the FSLE-3D defined filament for the two events shown in Figure \ref{fig:traj_y_fsle}. Panel (a) corresponds to the first event, and panel (b) to the second event. Each colored line represents the distance of an individual particle to the filament as a function of time.}
	%as influenced by the local FSLE structure.
	\label{fig:traj_distance}
\end{figure}

In both events, particle trajectories are strongly constrained by the FSLE-3D structures, which act as attracting material lines that organize horizontal transport \citep{Haller2015}. The stronger FSLE-3D values observed in the first event are consistent with an enhanced particle aggregation.
%whereas the weaker FSLE signal in the second event is associated with a less compact clustering. 
%In addition, secondary FSLE patches of lower intensity do not show a comparable ability to trap particles, suggesting that only the main filament cores dominate the transport dynamics. 
The figure provides evidence of the expected fact that FSLE-3D filaments act as transport organizers, controlling horizontal particle pathways. In the following we focus on the vertical dynamics.

Ten representative particles within each of the two sets plotted in Fig. \ref{fig:traj_y_fsle} are highlighted in red to track their evolution in detail. These particles were deliberately selected to be at different depths within the water column at the beginning of the tracking periods (27 February at 6:30, and 19 March at 21:30) to assess whether particles initially distributed vertically experience a similar dynamical evolution. %They correspond to sets of particles that were released at different heights. 
Although only a limited number of particles are explicitly tracked, their behavior is representative of the Lagrangian flow dynamics. 

Figure \ref{fig:traj_vertical} shows the temporal evolution in depth for each of the ten tracked particles for both filaments. Figure \ref{fig:traj_vertical}a corresponds to the first event, and Figure \ref{fig:traj_vertical}b corresponds to the second event. Their trajectories show a consistent behavior and a coherent dynamical response consistent with the converging behavior shown in Figure \ref{fig:traj_y_fsle}, indicating that the influence of the attracting structure extends throughout the sampled portion of the water column. In particular, all particles exhibit a generally-downward motion over the time interval associated with the presence  of the filament. In both cases, the depth–time representation shows that particle descent occurs when approaching the filament vicinity. This synchronization suggests a direct
coupling between the horizontal organization imposed by the FSLE-3D structures and the vertical particle dynamics. The filaments therefore act as pathways that not only organize horizontal transport but also guide particles downward in a coordinated manner.

Particles tend to descend in a parallel fashion, maintaining their relative separation while moving downwards. The total vertical displacement reaches tens of meters over timescales of one to two days, which is consistent with the depth range over which the filaments exhibit vertical coherence. This behavior is also in agreement with the Eulerian velocity cross-sections previously analyzed.

Superimposed on this overall descent, the particle trajectories exhibit vertical oscillations that may arise from the presence of internal waves, either of topographic origin or associated with internal tides, as observed in \cite{capo2022coherent}. However, the dominant transport mechanism, the persistent downward flow, is controlled by the filamentary structure.

The attracting character of the FSLE-3D structure is further seen in Figure \ref{fig:traj_distance}, where the temporal evolution of the horizontal distance between each particle and the corresponding surface FSLE-3D filament for the two events is analyzed. We recall that, because of the near-vertical nature of the three-dimensional structures identified, horizontal distance to the location of the surface filament gives a good approximation to the horizontal distance to the structure at each depth. Panel (a) corresponds to the first event, while panel (b) to the second. Although the selected particles were initially distributed at varying distances from the filament, their dynamics reveals a convergence toward the FSLE-3D defined structure. For both events, particles that are initially located several kilometers away from the filament undergo a rapid convergence (faster in the case of the stronger February filament than in the case of the March one) toward the filament core within a timescale of a few hours. As expected,  this convergence phase is more evident around the time of maximum FSLE-3D intensity.
Once near the filament, most particles remain close to it, and experience a downward motion. 
%Even particles relatively far from the filament experience with this downward motion, indicating that the dynamical influence of the identified structure extends beyond the filament core. 

These results combined with the vertical evolution in Figure \ref{fig:traj_vertical} show a  clear coupling  between horizontal and vertical dynamics. The rapid horizontal convergence toward the filament coincides with the strongest downward motion, while the period during which particles remain close to the filament corresponds to the interval of maximum vertical displacement. At later times, as particles move away from the filament (or the filament weakens or disappears),  vertical positions tend to stabilize. 

It is worth noting that FSLE, like  finite-time Lyapunov exponents, may quantify changes in the area of horizontal f luid patches, which are related to horizontal divergence and, because of fluid incompressibility, to vertical dynamics. But FSLE will also react to fluid-patch deformations due to strain. For instance, purely horizontal geostrophic  flows can exhibit strong FSLE signals associated with strain, despite having no associated vertical transport.

\section{Conclusions}
\label{Sec:summary}

In this study we have examined the three-dimensional transport properties associated with two selected coherent structures that extend vertically below surface filaments South of the Ibiza Channel, by combining FSLE diagnostics with ensemble Lagrangian trajectories. Our FSLE methodology was explicitly tuned to target submesoscale dispersion and convergence, but the structures identified extend beyond the submesoscale range, revealing the strong interplay between mesoscale and submesoscale features in determining fluid transport. 

Two-dimensional FSLE-2D fields at successive depths demonstrate that these structures are not limited to the surface but extend downward to several tens of meters, although the intensity of the FSLE-2D decreases with depth. This reduction shows a weakening of horizontal mixing, so that the role of filaments in organizing transport decreases as they move away from the surface. Our results are an example where the horizontal stirring structures identified at the surface extend into the ocean interior while preserving their geometrical organization.

The FSLE-3D calculations give results very similar to the FSLE-2D, confirming that at the scales analyzed here interparticle dispersion occurs mainly on the horizontal. 3D particle trajectories show systematic convergence toward the attracting filament cores, followed by a coherent downward displacement. The analysis of particle distances to the filaments quantifies this process. Particles rapidly converge toward the filament  within a few hours and remain close to it for a finite time, after which they gradually disperse as the filament weakens. 
%Even particles that do not remain in the close proximity to the filament exhibit a synchronized downward motion. This indicates that the dynamical influence of the filament extends beyond its core and affects a broader surrounding region. 
The vertical motion is of the order of tens of meters over timescales of days.

This descent is associated with the period of filament intensification, showing a coupling between horizontal convergence and vertical transport. Although oscillatory components appear also on the vertical particle motion, these are superimposed on a dominant downward drift, indicating that the 
filament-driven transport controls the vertical displacement.

Our results show that, in the two events examined here, FSLE-3D filaments act as three-dimensional transport structures that couple horizontal convergence with vertically coherent motion. The behavior is consistent with subduction-like processes associated with frontal dynamics. 

While this study focuses on two specific events, the similar behavior observed in both cases suggests that the coupling between horizontal convergence and downward transport may represent a recurring mechanism associated with coherent filamentary structures. However, the limited number of events considered here does not allow its generality to be established. Future work should therefore extend these diagnostics to a broader set of coherent structures, oceanic regions, and environmental conditions in order to assess the robustness and prevalence of this behavior.

A more complete understanding of these mechanisms could improve our ability to model 
three-dimensional transport in dynamically active regions, while potentially providing a pathway to infer aspects of its three-dimensional structure from remote sensing observations at the surface.

\section*{Author Contributions}
\noindent Conceptualization – SC, EHG and CL. 
Data curation – SC and IHC.  
Formal analysis – SC.   
Funding acquisition – AO, EHG and CL. 
Investigation – SC. 
Methodology – SC, IHC, EHG and CL.  
Project administration – AO, EHG and CL. 
Resources – SC, IHC and AO. 
Software – SC and IHC. 
Supervision – EHG and CL. 
Validation – SC.   
Visualization – SC. 
Writing – SC, EHG and CL.
Editing - SC, IHC, AO, EHG and CL.

\section*{Competing Interests}
None of the authors has any competing interests.
\section*{Declaration of generative AI and AI-assisted technologies in the manuscript preparation process.}
During the preparation of this work, the author(s) used OpenAI and Microsoft Copilot to check the text for coherence, clarity, and typographical errors. After using these tools, the author(s) reviewed and edited the content as needed and take(s) full responsibility for the content of the published article.
\section*{Acknowledgements}
We gratefully acknowledge financial support by grants PID2021-123352OB-C31, PID2021-123352OB-C32 and PID2023-153236NA-I00 funded by MICIU/AEI/10.13039/501100011033  and FEDER, "Una manera de hacer Europa". E.H-G also acknowledges the Mar\'{\i}a de Maeztu project CEX2021-001164-M funded by\\
MICIU/AEI/10.13039/501100011033.

%% REFERENCES

%% The reference list is compiled as follows:
\newpage
\bibliographystyle{elsarticle-harv}
\bibliography{biblio.bib}

%% LITERATURE CITATIONS
%%
%% command                        & example result
%% \citet{jones90}|               & Jones et al. (1990)
%% \citep{jones90}|               & (Jones et al., 1990)
%% \citep{jones90,jones93}|       & (Jones et al., 1990, 1993)
%% \citep[p.~32]{jones90}|        & (Jones et al., 1990, p.~32)
%% \citep[e.g.,][]{jones90}|      & (e.g., Jones et al., 1990)
%% \citep[e.g.,][p.~32]{jones90}| & (e.g., Jones et al., 1990, p.~32)
%% \citeauthor{jones90}|          & Jones et al.
%% \citeyear{jones90}|            & 1990

\end{document}